\documentclass[aps,prl,twocolumn,nofootinbib,notitlepage,superscriptaddress,aps,pra]{revtex4-2}
\usepackage{float}
\usepackage{amsmath}
\usepackage{amssymb}
\usepackage{graphicx}
\usepackage{soul}
\usepackage{physics}
\usepackage{color}
\usepackage{mathtools}
\usepackage[colorlinks=true,linkcolor=black,anchorcolor=black,citecolor=blue]{hyperref}
\usepackage{cleveref}
\usepackage{tabularx}

\begin{document}
\title{{\normalsize Planckian bound on IR/UV mixing from cold-atom interferometry}
}
\author{G. Amelino-Camelia}
\affiliation{Dipartimento di Fisica Ettore Pancini, Universit\`a di Napoli ``Federico II", Complesso Univ. Monte S. Angelo, I-80126 Napoli, Italy}
\affiliation{INFN, Sezione di Napoli, Complesso Univ. Monte S. Angelo, I-80126 Napoli, Italy}
\author{G. Fabiano}
\affiliation{Physics Division, Lawrence Berkeley National Laboratory, Berkeley, CA}\affiliation{Department of Physics, University of California, Berkeley, CA}\affiliation{Centro Ricerche Enrico Fermi, I-00184 Rome, Italy}
\author{D. Frattulillo}
\affiliation{INFN, Sezione di Napoli, Complesso Univ. Monte S. Angelo, I-80126 Napoli, Italy}
\author{F. Mercati}
\affiliation{Departamento de F\'isica, Universidad de Burgos, 09001 Burgos, Spain}
\begin{abstract}
IR/UV mixing (a mechanism causing  ultraviolet quantum-gravity effects to manifest themselves also in a far-infrared regime) is a rare case of feature found in several 
approaches to the quantum-gravity problem. 
We here derive the implications for
"soft" IR/UV mixing (corrections to the dispersion relation that are linear in momentum)
of some recent cold-atom-interferometry measurements. For both signs of the 
IR/UV-mixing correction term we establish bounds on the characteristic length scale which reach the Planck-length milestone. Intriguingly, for values of the characteristic scale of about half the Planck length we find that 
IR/UV mixing provides a solution for a puzzling discrepancy between Cesium-based and Rubidium-based atom-interferometric measurements of the fine structure constant. 
\end{abstract}
\maketitle

%

Quantum gravity presents itself as an ultraviolet problem: the tension between general relativity and quantum mechanics is increasingly severe as the energy of a microscopic particle increases.
However,  several approaches to the quantum-gravity problem predict that there should be some infrared counterparts to
the novelties introduced in the ultraviolet regime, a mechanism known as IR/UV (infrared/ultraviolet) mixing.
This mechanism might be as structural to the quantum-gravity problem as Hawking radiation,
and indeed the Bekenstein-Hawking entropy-area relation reflects an aspect
of IR/UV mixing \cite{Cohen:1998zx}.

The relevant phenomenology
has mainly analyzed
the implications of
IR/UV mixing for the dependence of the energy of a massive particle 
on its spatial momentum in the infrared regime such that 
the spatial momentum is  smaller than the mass. When spacetime quantization is formalized in terms
of noncommutativity of coordinates one can have (depending on the assumed form of coordinate noncommutativity) either "hard
IR/UV mixing", with a correction to the energy of the particle going like the inverse square power of spatial momentum \cite{Minwalla:1999px,Khoze:2000sy},
or "soft
IR/UV mixing", 
with correction to the energy depending linearly on spatial momentum \cite{Matusis:2000jf,Szabo:2001kg}.
Also some quantum-spacetime models based on discreteness predict 
\cite{Alfaro:1999wd} an infrared correction to the energy of massive particles depending linearly on spatial momentum.

We here focus on soft IR/UV mixing
and set up the analysis in terms of a relationship between energy ($E$) mass ($m$) and spatial momentum ($p$) given by \cite{Alfaro:1999wd}

\vspace{-0.45cm}

\begin{equation}
\label{eq:mdr1alternate1}
    E=m +\frac{p^2}{2m}+ \ell \, {m \, p \, }
\end{equation}
where $\ell$ is a phenomenological parameter, which could have either sign, with dimensions of length (in units such that the reduced Planck constant and the speed of light are set to 1). It is expected \cite{Alfaro:1999wd} that $|\ell|$
should be close to the Planck length $L_P$ ($\simeq 1.616 \cdot 10^{-35} m$).

A possible role for this IR/UV-mixing scenario in cold-atom interferometry was already considered in previous studies  (see, {\it e.g.}, Refs.~\cite{Amelino-Camelia:2009wvc,Mercati:2010au,Briscese:2012ip,Castellanos:2012cs,Carmona:2013fwa,Albrecht:2014sxa,Plato:2016azz,Freidel:2021wpl}): the relevant interferometers involve several stages in which atoms with  momentum much smaller than their mass  get their momentum changed through interactions with photons, and clearly the IR/UV-mixing correction term in Eq.(\ref{eq:mdr1alternate1})  affects the kinematics of atom-photon interactions. Our main objective is to show that results obtained with the latest generation of cold-atom interferometers can be used to set bounds on $\ell$ that reach the milestone $|\ell|<L_P$.

We start by considering the latest cold-atom-interferometry measurements of the fine structure constant $\alpha$
also taking into account  
measurements of $\alpha$  based on the anomalous magnetic moment (g-2) of the electron.

\vspace{-1.61cm}

\begin{figure}[h!]
    \centering
\includegraphics[scale=0.7]{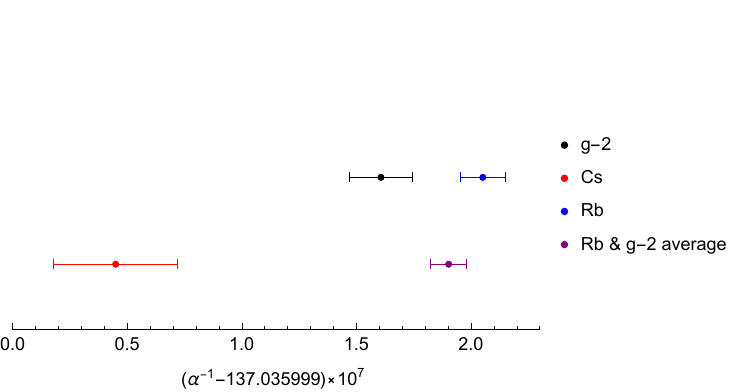}
    \caption{The black data point is for the present status  of $\alpha$  measurements using the electron g-2 \cite{Fan:2022eto,newnumbers}($\alpha_{g-2}^{-1}=137.035999160(14)$). The red and blue data points are for the latest atom-interferometry measurements of $\alpha$ using, respectively, Cesium \cite{parker2018measurement,newnumbers}
($\alpha_{Cs}^{-1}=137.035999045(27)$) and Rubidium \cite{Morel:2020dww,newnumbers} ($\alpha_{Rb}^{-1}=137.0359992052(97)$). The  weighted average of the g-2 and Rubidium measurements
is in violet.}    \label{fig:finestruct}
\end{figure}

As shown in  our \Cref{fig:finestruct} the  measurement of the fine structure constant $\alpha$ using  an interferometric setup  with Rubidium atoms \cite{Morel:2020dww} (see Figure 2)
is in rather good agreement with the measurement of $\alpha$  based on  the electron g-2 \cite{Fan:2022eto}, but there is a  sizable discrepancy between those results and  
the  measurement of $\alpha$ using  an interferometric setup  with Cesium atoms \cite{parker2018measurement} (see Figure 3).
[Note that, as commonly done, the schematic  description of atom interferometers given in Figures 2 and 3 neglect the effects of gravity, which bend the space-time trajectories of atomic beams.]

\begin{figure}[h!]
    \centering
    \includegraphics[scale=0.57]{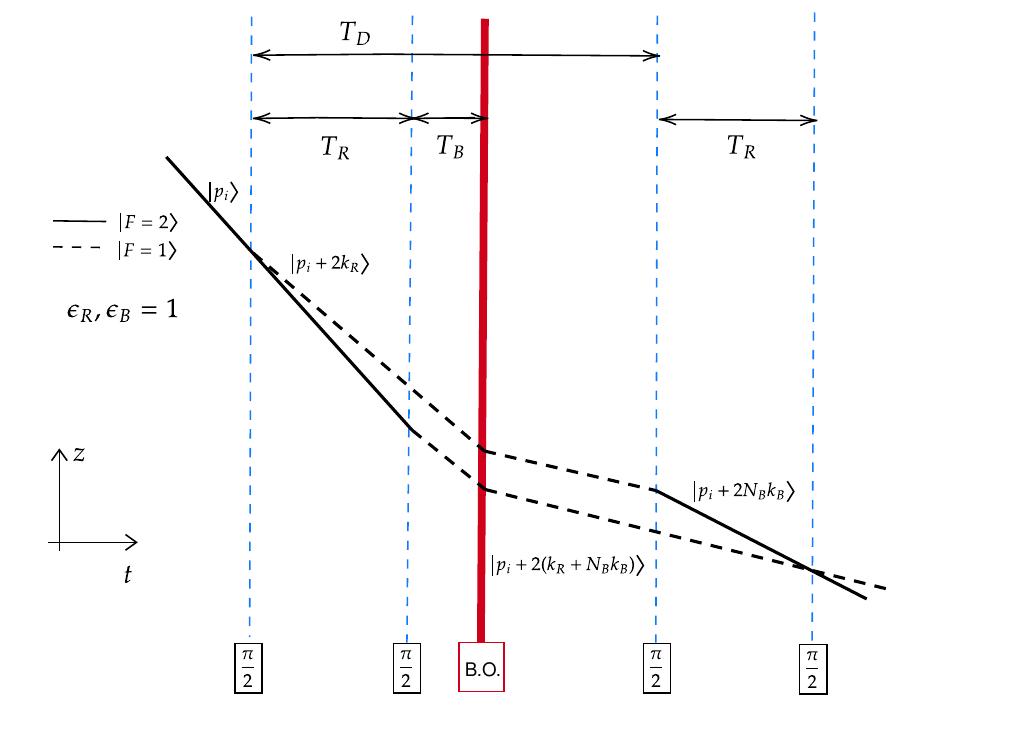}
    \caption{Schematic representation of one of the four configurations ($\epsilon_R=\epsilon_B=1$, see \Cref{subsec:appendixrubidio})  of the Rubidium interferometer of Ref.\cite{Morel:2020dww}, a Ramsey-Bordé in differential velocity sensor configuration.
The black lines, labeled by momentum, represent trajectories of atomic beams,
dashed black lines for atoms in  $\ket{F=1}$ internal state, thick black lines for atoms in  $\ket{F=2}$ internal state. The beam starts with momentum $p_i$, and additional momentum is imparted by Raman  transitions (dashed blue lines) and Bloch oscillations (thick red line).  $T_R,T_B,T_D$ are the time intervals  between interactions.}
    \label{fig:rubidio1}
\end{figure}

\begin{figure}
    \centering
    \includegraphics[scale=0.41]{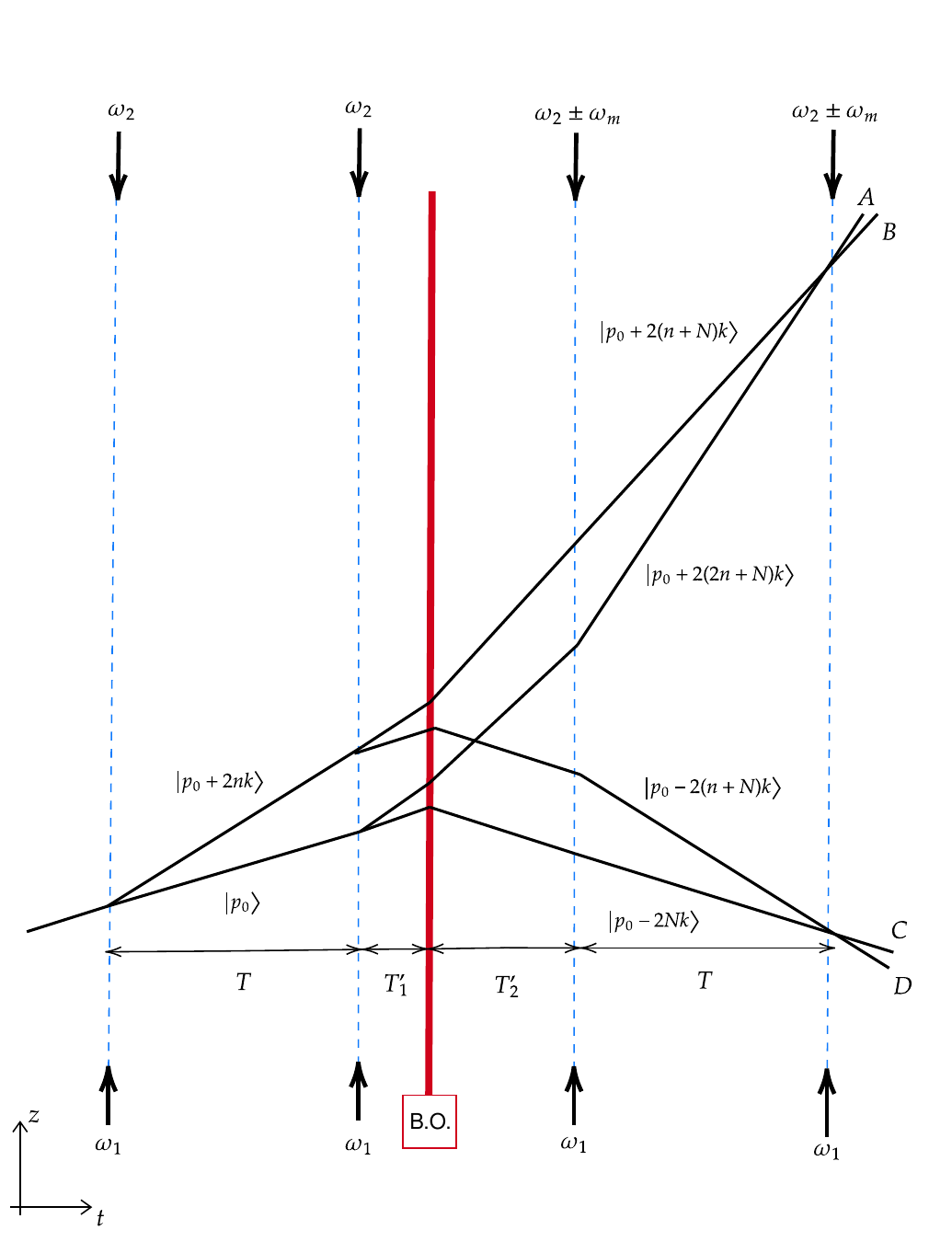}
    \caption{
Schematic representation of the simultaneous conjugate Ramsey-Bordé Cesium interferometer of Ref.\cite{parker2018measurement}.  Thick black lines represent  trajectories of atomic beams, labeled by momentum. The beam starts with momentum $p_0$, and additional momentum is imparted (without affecting internal energy) by means of Bragg diffraction (dashed blue lines) and Bloch oscillations (thick red line).
$T,T'_1,T'_2$ are the time intervals between   interactions.  $\omega_1$, $\omega_2$ are the frequencies of the lasers employed for the first two Bragg diffractions, tuned to impart $\pm n k$ momentum, where $k=(\omega_1+\omega_2)/2$. The Bloch oscillations impart $N k$ to beams A and B and $-N k$ to beams C and D. The frequency $\omega_2$ in the third and fourth Bragg diffractions is corrected by an amount $\omega_m$  for arms C and D and by $- \omega_m$ for arms A and B (see \Cref{subsec:appendixcesio}).}
    \label{fig:cesio1}
\end{figure}

The discrepancy highlighted in our \Cref{fig:finestruct}
is perceived as a major puzzle for fundamental physics (see, {\it} e.g., Ref.~\cite{newnumbers} and references therein).
From the perspective of IR/UV mixing it is natural to investigate whether the content of  \Cref{fig:finestruct}
 could change significantly by allowing for the IR/UV-mixing correction term of Eq.~(\ref{eq:mdr1alternate1}).
Measurements of $\alpha$ based on g-2 of the electron involve particles with spatial momentum larger than their mass, and therefore   the IR/UV correction has negligible effects there.
Instead, the IR/UV correction has tangible effects
(even for $|\ell| \sim L_P$) in several stages of the interferometric setups of the
Rubidium-based and
of the Cesium-based $\alpha$ measurements; however, in the interferometric setup of the Rubidium-based measurement of Ref.\cite{Morel:2020dww}, as shown in \Cref{subsec:appendixrubidio}, there is a large cancellation of IR/UV-mixing correction terms resulting in no net effect.
For the interferometric setup of the Cesium-based measurement of Ref. \cite{parker2018measurement}  we find, as shown in \Cref{subsec:appendixcesio}, 
that the overall IR/UV-mixing correction can be cast in the form of a shift in the determination of the fine structure constant, given by
$$\alpha = \alpha_0 \left(1 - \ell \cdot 1.08\cdot 10^{26} \, \text{m}^{-1}\right)$$
where $\alpha_0$ is the estimate of the fine structure constant  reported in Ref. \cite{parker2018measurement}, assuming that there is no IR/UV mixing.

In light of these findings the content of our
\Cref{fig:finestruct}
can be viewed as the basis for a measurement of
$\ell$:
combining the measurements of the fine structure constant based on Rubidium atoms and on the electron g-2, reported in \cite{Morel:2020dww,newnumbers} and
\cite{Fan:2022eto,newnumbers} respectively  (violet point in \Cref{fig:finestruct}), we get the present best estimate of $\alpha$ not affected by $\ell$
and then requesting consistency with the $\alpha$ measurement of Ref. \cite{parker2018measurement} one finds
\begin{equation}
\ell = (9.8 \pm 1.9)\cdot 10^{-36} \,\text{m} = (0.60 \pm 0.12) L_P
\label{ellalpha}
\end{equation}
It is intriguing that for a value of $\ell$ of about half the Planck length, which is still  within the range of values of $\ell$ considered plausible in the quantum-gravity literature \cite{Addazi:2021xuf}, IR/UV mixing can solve the discrepancy highlighted in our  Fig. \ref{fig:finestruct}.
From a more conservative perspective it is rather significant that our result (\ref{ellalpha})
sets the bound $\ell < L_P$ with high confidence ($99.96\%$) for positive $\ell$,
and establishes very robustly $- \ell < L_P $ for negative $\ell$.

Obtaining 
our result (\ref{ellalpha})
prompted us to do an extensive literature search for other measurement results which might be relevant: reaching Planckian sensitivity is considered a very significant milestone in quantum-gravity research, and frontier experiments are inevitably subject to the risk of unnoticed systematic errors, so it is undesirable  to set a Planckian bound on the basis of a single measurement.
We did find a potentially relevant measurement, also based on cold-atom interferometry: the test of the equivalence principle (EP)
reported in Ref.\cite{Asenbaum:2020era}, which measured 
 the relative acceleration of freely
falling clouds of atoms of $^{85}$Rb and $^{87}$Rb with a dual-species atom interferometer (see Figure 4), finding for the Eötvös parameter\footnote{In tests of the EP it is standard to measure the Eötvös parameter $\eta$, which is the relative acceleration of two test masses
divided by the average acceleration between the test
masses and the gravitational source.}
$\eta=\big[1.6\pm 1.8 (\text{stat})\pm 3.4 (\text{sys})\big]\cdot 10^{-12}$. 

\begin{figure}[h!]
    \centering
\includegraphics[scale=0.55]{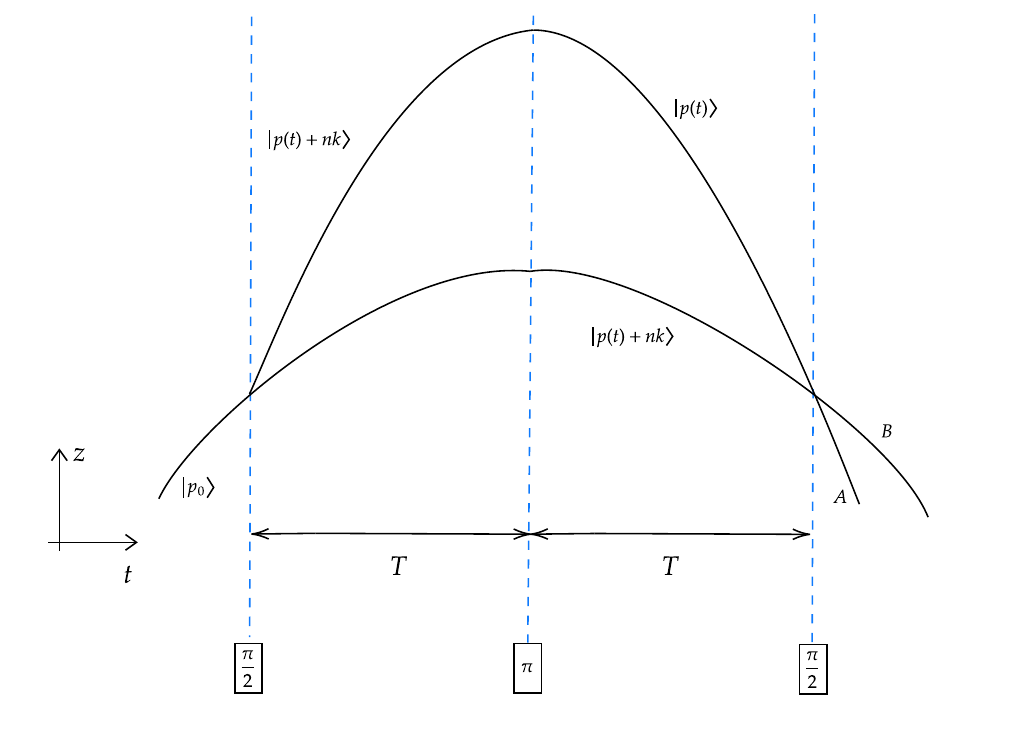}
    \caption{Schematic representation of the setup of the dual species atom interferometer of Ref.~\cite{Asenbaum:2020era} using $^{85}$Rb and $^{87}$Rb atoms. For clarity, only one of the Rubidium species is shown. The black lines labeled by momentum represent the trajectories of atomic beams $A$ and $B$. The beam starts with momentum $p_0$ and additional momentum $\pm n k$ is imparted by Bragg diffraction (dashed blue lines), where $k$ is the wave number of the laser. The change of the momentum due to gravity is denoted by $p(t)$ and $T$ denotes the time interval between interactions (see \Cref{sec:epinterferometer}).}    \label{fig:epinterforemeter}
\end{figure}

As shown in \Cref{sec:epinterferometer},
we find
that in presence of the IR/UV-mixing correction term of Eq.~(\ref{eq:mdr1alternate1}) the analysis reported in Ref.\cite{Asenbaum:2020era} would change significantly, resulting in a shift in the determination of the Eötvös parameter, given by
$$\eta = \eta_0 \left(1 + \ell  \cdot 3.77\cdot 10^{35} \, \text{m}^{-1}\right)$$
where $\eta_0$ is the estimate of the Eötvös parameter reported in Ref.~ \cite{Asenbaum:2020era}, assuming that there is no IR/UV mixing.

Making the robust assumption that the EP holds at least at the level\footnote{The measurement result
$\eta=\big[1.5\pm 2.3 (\text{stat})\pm 1.5 (\text{sys})\big]\cdot 10^{-15}$ for the
Eötvös parameter 
was obtained  
by comparing the gravitational accelerations of two macroscopic test masses of different composition \cite{MICROSCOPE:2022doy}. In order to derive our result (\ref{ellEP}) we assume that the EP holds at the same level for 
$^{85}$Rb and $^{87}$Rb atoms. Future studies contemplating combined effects of IR/UV mixing and of enhanced violations of the EP for atomic systems, could use
the experimental results reported in Ref. \cite{Asenbaum:2020era} to set combined bounds on $\ell$ and on the value of $\eta$ for  $^{85}$Rb/$^{87}$Rb atoms.
} $|\eta|<10^{-14}$ \cite{MICROSCOPE:2022doy}, 
one then finds that the results reported in Ref.\cite{Asenbaum:2020era} amount to a measurement of the characteristic $\ell$ of IR/UV mixing: 
\begin{equation}
\ell = (- 2.6 \pm 6.4 )\cdot 10^{-36} \,\text{m} = (-0.16 \pm 0.40) L_P
\label{ellEP}
\end{equation}
We therefore have our sought second measurement establishing Planckian bounds on the IR/UV mixing scenario of Eq.(\ref{eq:mdr1alternate1}): our result (\ref{ellEP}) 
sets the bound $\ell < L_P$ with high confidence ($99.5\%$) for positive $\ell$,
and also establishes $- \ell < L_P $ with high confidence ($97.3\%$) for negative $\ell$.

Within uncertianties, our result (\ref{ellEP}) is consistent with $\ell =0$ but is also consistent with the estimate of $\ell$ given by our result (\ref{ellalpha}), and therefore does not weaken the prospects that IR/UV mixing might solve the discrepancy highlighted in our  Fig. \ref{fig:finestruct}.
At the current pace of improvement of the accuracy of cold-atom interferometers it should not take long to investigate this exciting possibility.  For example, in order to settle this issue it might suffice to improve  
by a factor of 3 the accuracy of an interferometer like the one of 
Ref.~\cite{Asenbaum:2020era}.
\section*{Acknowledgements}
We are grateful to P. Asenbaum, M. A. Kasevich, H. Müller, C. Overstreet, G. Tino and W. Zhong for clarifications on some of the interferometers relevant for our study. G.A.-C. is grateful for financial support by the Programme STAR Plus, funded by Federico II University and Compagnia di San Paolo.
G.F.'s work on this project was supported by ``The Foundation Blanceflor".
F.M. acknowledges support by the Agencia Estatal de Investigación (Spain)
under grants CNS2023-143760 and PID2023-148373NB-I00 funded by MCIN/AEI/10.13039/501100011033/FEDER – UE,
and by the Q-CAYLE Project funded by the Regional Government of Castilla y León (Junta de Castilla y León) and by the Ministry of Science and Innovation MICIN through NextGenerationEU (PRTR C17.I1).
This work also benefited from the activities of the European Union COST Action CA23130 \emph{Bridging high and low energies in search of quantum gravity}.

\appendix
\section{Phase shift calculations}
\label{sec:appendix}
In this appendix we review the main steps for the calculation of the total phase shift for the atom interferometry configurations analyzed in \cite{parker2018measurement,Morel:2020dww,Asenbaum:2020era}. First, we present the general formalism, following \cite{overstreetth,Storey:1994oka}, and then apply it to the specific experimental configurations, following \cite{zhongth,morelth,overstreetth}. The contributions to the total phase shift include:

\begin{itemize}
    \item Propagation phase: In the path  integral approach proposed in \cite{Storey:1994oka}, the propagation phase  is written in terms of the action
\begin{equation}
    {S(\Gamma)}=\int_\Gamma\mathcal{L} \, dt \, ,
\end{equation}
with $\mathcal{L}$ being the Lagrangian and $\Gamma$ being the path along which the action is evaluated. For atom interferometers, it turns out that $S(\Gamma)\gg 1$ \cite{zhongth,morelth}, so for our purposes the action can be evaluated along the classical trajectory $\Gamma_{cl}$ which extremizes it. The propagation phase is then defined as the difference in the action computed along the classical trajectory of each interferometer arm, as follows
\begin{equation}
\label{eq:propphase}
\begin{aligned}
    \phi^p=&\int_{t_0}^{t_f}\mathcal{L}(\vec{x}_1(t),\dot{\vec{x}}_1(t),t) \, dt +\\
    -& \int_{t_0}^{t_f}\mathcal{L}(\vec{x}_2(t),\dot{\vec{x}}_2(t),t) \, dt \, .
\end{aligned}
\end{equation}
In the above, $t_0$ is the time of the initial beamsplitter, $t_f$ is the time of the final beamsplitter and $\vec{x}_i(t)$ is the coordinate of the classical trajectory along the $i$-th arm of the interferometer.
\item Matter-radiation interaction phase: this phase contribution arises whenever an atom-photon interaction occurs. For an interaction occuring at the $i$-th arm of the interferometer, it is given by \cite{Storey:1994oka}
\begin{equation}
\label{eq:rmphase}
    \phi^{{int}}=\pm \left(\Vec{k}\cdot\Vec{x}_i(t_i)-\omega \, t_i\right)
\end{equation}
where the $+/-$ signs indicate whether the photon has been absorbed or emitted, respectively. The pair $(\Vec{k},\omega)$ denotes the wave vector and frequency of the absorbed/emitted photon while  $(\Vec{x}_i(t_i),t_i)$ are the space-time coordinates indicating where and when the interaction takes place. When calculating the total phase shift, all the interaction contributions of the lower arm of the interferometer are subtracted from the ones of the upper arm, yielding the total matter-radiation interaction phase.

\item Separation phase: In the semi-classical approximation, the classical trajectories need not overlap at the time of the final beamsplitter pulse, giving rise to an open interferometer. This issue introduces an ambiguity in calculating the total phase, since the result will be different if the final beamsplitter interaction is considered for the upper or lower interferometer. To resolve the issue, a separation phase is introduced, which averages over the final momenta of the beams at the output ports and takes into account the distance between the arms at the time of the final beamsplitter. The separation phase is given by
\begin{equation}
\label{eq:sepphase}
    \phi^{sep}=\left(\frac{\vec{p_1}(t_f)+\vec{p_2}(t_f)}{2}\right)\cdot\bigg(\vec{x_2}(t_f)-\vec{x_1}(t_f)\bigg)
\end{equation}
\end{itemize}
The total phase for the atom interferometer is obtained by summing the propagation phase, the matter-radiation interaction phase averaged over the two arms of the interferometer at the two different output ports and the separation phase. In formulas
\begin{equation}
    \phi^{tot}=\phi^p+\sum_{interactions}\bar{\phi}^{int}+\phi^{sep} \, ,
\end{equation}
where the sum takes into account all of the matter-radiation interactions in the interferometric sequence and $\bar{\phi}^{int}$ indicates that the interaction phase at the two output ports is averaged over the two atomic wave packets.

The expression for the Lagrangian that takes into account the modification to the dispersion relation as in \eqref{eq:mdr1alternate1} and the gravitational potential is given by
\begin{equation}
\label{eq:lagrangian}
    \mathcal{L}=-m+\frac{mv^2}{2}-\ell {m^2 v}-m g z \, .
\end{equation}
The experimental sequence we are considering can be described with good approximation in 1 spatial dimension, which we will identify as the $z$-axis.
The relation between the spatial component of the momentum and velocity is given by
\begin{equation}
\label{eq:velocità}
    v^z=\frac{\partial E}{\partial p^z}=\frac{p^z}{m}+ \ell m \chi(p^z)  \, ,
\end{equation}
where $E$ is given by \eqref{eq:mdr1alternate1}, $\chi(p^z)=1$ if $p^z>0$ and $\chi(p^z)=-1$ if $p^z<0$.
The trajectory can be obtained by integrating the above expression
\begin{equation}
\label{eq:trajectory}
\begin{aligned}
    &z(t)=\int_{t_0}^t v^z(t') \, dt'=\\
    &\int_{t_0}^t \left[\frac{p^z(t')}{m}+\ell m \chi(p^z(t')) \right]\, dt' \, ,
\end{aligned}
\end{equation}
with $p^z(t)=p_0^z-mgt$, where $t_0$ is such that $p^z(t_0)=p_0^z$.
Given the expressions \eqref{eq:lagrangian},\eqref{eq:velocità},\eqref{eq:trajectory}, we can calculate all of the main contributions to the interferometer phase.

\subsection{Rubidium atom interferometer measuring $\alpha$}
\label{subsec:appendixrubidio}
The measurement of the fine structure constant with Rubidium atom interferometry reported in \cite{Morel:2020dww} is based on the Ramsey-Bordé interferometer  in differential velocity sensor configuration \cite{morelth,BORDE198910} depicted in \Cref{fig:rubidio1}. The main idea of this setup is to measure the transferred velocity to the atoms via Bloch oscillations (indicated by the thick red line) in between two Ramsey sequences separated by a delay $T_D$. This velocity transfer is $\Delta v=2 N_B k_B/{m_{Rb}}$, where $N_B$ is the number of Bloch oscillations ($\sim 500$), $k_B$ is the Bloch beams wavevector and $m_{Rb}$ is the Rubidium atom mass. The Ramsey sequences, of duration $T_R$, are characterized by two Raman based beam splitters operating with lasers of wave number $k_R$. The beam splitters are indicated by the $\pi/2$ boxes in the sequence, along with the dahsed black lines. Additionally, in order to keep the Raman resonance condition along the interferometer, a ramp for each of the Ramsey sequences is needed and an additional frequency jump due to velocity transfer process has to be taken into account. This is obtained by experimentally allowing for a time dependence of the laser frequencies $\omega(t)$. We denote by $\delta \omega_R$ the difference between $\omega(T_D)$ and $\omega(0)$.
Effects due to gravity and due to light shifts (encoded in an additional phase $\phi_{LS}$) \cite{Morel:2020dww,morelth} can be eliminated by combining the signals obtained from a total of four different configurations in which Bloch accelerations are in opposite directions ($\epsilon_B=\pm 1$) and the directions of Raman transitions are inverted ($\epsilon_R=\pm 1$). In the undeformed case, a straightforward calculation of the propagation, interaction and separation phases introduced in \eqref{eq:propphase},\eqref{eq:rmphase},\eqref{eq:sepphase}, yields the total phase for each configuration

\begin{equation}
\begin{aligned}
\label{eq:faserubidio}\Phi(\epsilon_R,\epsilon_B)=&\phi_A(\epsilon_R,\epsilon_B)-\phi_B(\epsilon_R,\epsilon_B)=\\
=&2T_R\epsilon_R k_R\left( \frac{2 \epsilon_BN_B  k_B}{m_{Rb}}-gT_D\right) + \\
+&\phi_{LS}-T_R\delta \omega _R(\epsilon_R,\epsilon_B) \, ,
\end{aligned}
\end{equation}
where $\phi_A(\epsilon_R,\epsilon_B),\phi_B(\epsilon_R,\epsilon_B)$ are the accumulated phases in arms $A,B$ of the interferometer configuration in \Cref{fig:rubidio1}. 
In each case, $\delta\omega_R(\epsilon_R,\epsilon_B)$ is adjusted in order to obtain $\Phi(\epsilon_R,\epsilon_B)=0$. 
Averaging over the four combinations of $\epsilon_R,\epsilon_B$, one determines  $1/m_{Rb}$:
\begin{equation}\label{hrubidium}
    \left(\frac{1}{m_{Rb}}\right)=\frac{\sum_{\epsilon_R,\epsilon_B}\left|{\delta\omega_R(\epsilon_R,\epsilon_B)}\right|}{16N_Bk_Bk_R} \, .
\end{equation}
The measurement of the fine structure constant is then determined through the relation $\alpha^2=4\pi R_{\infty}\frac{m_{Rb}}{m_e}\frac{1}{m_{Rb}}$, where $R_{\infty}$ is the Rydberg constant and $m_e$ is the electron mass.
Upon introducing the modified dispersion relation \eqref{eq:mdr1alternate1}, the corrections to the phases accumulated in the two arms of the interferometer of \Cref{fig:rubidio1}, denoted by $\delta\phi_A(\epsilon_R,\epsilon_B)$ and $\delta\phi_B(\epsilon_R,\epsilon_B)$, are found to be equal to each other, and specifically

\begin{equation}
    \begin{aligned}\delta\phi_A(\epsilon_R,\epsilon_B)=&
    {\ell  m_{Rb}}\bigg(-\frac{m_{Rb}g}{2}(T_D+T_R)^2+\\
    -&2k_BNT_B\epsilon_B+k_R(T_D-T_R)\epsilon_R\bigg)\, ,
    \end{aligned}
\end{equation}
\begin{equation}
    \begin{aligned}\delta\phi_B(\epsilon_R,\epsilon_B)=&
    {\ell  m_{Rb}}\bigg(-\frac{m_{Rb}g}{2}(T_D+T_R)^2+\\
    -&2k_BNT_B\epsilon_B+k_R(T_D-T_R)\epsilon_R\bigg)\, ,
    \end{aligned}
\end{equation}
Therefore, the correction to the total phase for each configuration is $\delta\Phi(\epsilon_R,\epsilon_B)=\delta\phi_A(\epsilon_R,\epsilon_B)-\delta\phi_B(\epsilon_R,\epsilon_B)=0$ so that the determination of the fine structure constant using the interferometric setup of Ref. \cite{Morel:2020dww} is unchanged when introducing IR/UV mixing corrections.

\subsection{Cesium atom interferometer measuring $\alpha$}
\label{subsec:appendixcesio}
The measurement of the fine structure constant with Cesium atom interferometry reported in \cite{parker2018measurement} is based on the Ramsey-Bordé configuration \cite{BORDE198910,zhongth}. The idea is to use two simultaneous conjugate interferometers, in order to cancel out effects due to gravity in the calculation of the total phase, rendering the measurement more precise. The interferometric sequence is depicted in \Cref{fig:cesio1}. This particular setup employs Bragg diffraction as beam splitters (the light blue dashed lines in \Cref{fig:cesio1}), which has the advantage of changing the momentum of an atomic beam without changing the internal energy levels of the atoms. The effect of each beam splitter drives the atomic beam into a superposition of momentum states, which differ by multiples of $k$, where $k$ is the wave number of the lasers. In order to increase experimental sensitivity, a sequence of Bloch oscillations is also imparted to the atom beams to increase their momenta by $N k$, where $N$ is of order $10^2$ in the experiment. The wave packets of the two simultaneous conjugate interferometers are recombined at the last beam splitter and interference patterns are observed.
Notice that at the third and fourth beam splitters, the laser frequency is adjusted by a term $\pm \omega_m$, where the $+$ sign is for the CD interferometer while the $-$ sign is for the AB interferometer, driving the conjugate pairs further apart.
Without the IR/UV corrections, the formula for the total phase is given by

\begin{equation}
\label{eq:csord0}
\begin{aligned}
    \Phi=&(\phi_{A}-\phi_B)-(\phi_C-\phi_D)=\\
    =&2nT\left(\omega_m-\frac{4(n+N) k^2}{m_{Cs}}\right) \, ,
    \end{aligned}
\end{equation}
where $\phi_A,\phi_B,\phi_C,\phi_D$ are the accumulated phases in arms $A,B,C,D$ of the interferometer configuration in \Cref{fig:cesio1}, respectively, $n$ is the order of the Bragg diffraction, $T$ is the free fall time interval between the first two interactions and $m_{Cs}$ is the mass of the Cesium atom. The frequency $\omega_m$ is adjusted in the experiment so that $\Phi=0$. This yields an indirect measurement of $1/{m_{Cs}}$ in terms of the directly measured $\omega_m$:
\begin{equation}\label{eq: h/m}
    \left(\frac{1}{m_{Cs}}\right)=\frac{ \omega_m}{4(n+N)k^2} \, ,
\end{equation}
In turn, this yields a measurement of $\alpha$ using the relation $\alpha^2=4\pi R_{\infty}\frac{m_{Cs}}{m_e}\frac{1}{m_{Cs}}$.

The introduction of the modified dispersion relation \eqref{eq:mdr1alternate1} yields corrections to the total phase \eqref{eq:csord0}. Denoting by $\delta\phi_I$, with $I=A,B,C,D$ , the IR/UV mixing corrections to the phases accumulated in the four arms of the interferometric scheme of \Cref{fig:cesio1}, we have

\begin{equation}
    \begin{aligned}\delta\phi_A=&\frac{\ell  m_{Cs}}{4}\bigg(gm_{Cs}(2T+T'_1+T'_2)^2+\\
    +&2 k (n(2T+T'_1+T'_2)+2(T+T'_1)\bigg) \, ,
    \end{aligned}
\end{equation}
\begin{equation}
    \begin{aligned}\delta\phi_B=&\frac{ \ell m_{Cs}}{4}\bigg(gm_{Cs}(2T+T'_1+T'_2)^2+\\
    +&2 k (n(2T+T'_1+T'_2)+2(T+T'_1)\bigg) \, ,
    \end{aligned}
\end{equation}

\begin{equation}
\begin{aligned}
    \delta\phi_C=&\frac{ \ell m_{Cs}}{2}\bigg(-2 k\big(n(T'_1-T'_2)+\\
    &+2N(T'_1+T)\big)+\\
&+gm_{Cs}\big(2(T+T'_1)^2-(T'_1-T'_2)^2\big)\bigg) \, ,\\
    \end{aligned}
\end{equation}

\begin{equation}
\begin{aligned}
    \delta\phi_D=&\frac{\ell  m_{Cs}}{2}\bigg(-2 k\big(n(4T+T'_1-T'_2)+\\
    &+2N(T'_1+T)\big)+\\
    &+gm_{Cs}\big(2(T+T'_1)^2-(T'_1-T'_2)^2\big)\bigg) \, ,\\
    \end{aligned}
\end{equation}
where $T'_1,T'_2$ are time intervals shown in \Cref{fig:cesio1}.
Following the procedure outlined for the undeformed case, from the expression of the total phase it is possible to extract a measurement of $\alpha_{Cs}$, which in terms of the undeformed measurement, which we denote by $\alpha_{Cs,0}$, is given by the relation
\begin{equation}
\label{eq:cscorr}
     \alpha_{Cs}=\alpha_{Cs,0}\left(1-\ell \frac{m_{Cs}^2}{4 k(n+N)}\right) \, .
\end{equation}

\subsection{Rubidium atom interferometer testing EP}
\label{sec:epinterferometer}
The test of the equivalence principle reported in \cite{Asenbaum:2020era} is based on the interferometric scheme depicted in \Cref{fig:epinterforemeter}, apt to measure the relative acceleration between ${}^{85}$Rb and ${}^{87}$Rb atoms. The idea is that the phase accumulated by each isotope is proportional to its gravitational acceleration, so the difference between these two phases becomes proportional to the E\"{o}tv\"{o}s parameter. The interferometer sequence is as follows. The atoms are launched vertically and at $t=0$ a first beamsplitter, implemented in terms of Bragg diffraction, drives the atomic beam in a superposition of momentum states which differ by multiples $k$, where $k$ is the wave vector of the laser. At $t=T$, a second Bragg diffraction acts as a mirror and reverses the momentum kicks of the two atomic beams, driving them towards each other. At $t=2T$, a final beamsplitter recombines the atomic beams and interference patterns are observed. For each species, the accumulated phase is given by
\begin{equation}
    \Phi^{85(87)}=\phi^{85(87)}_A-\phi^{85(87)}_B=-nkg_{85(87)}T \, ,
\end{equation}
where $\phi^{85(87)}_A$ and $\phi^{85(87)}_B$ are the accumulated phases for ${}^{85(87)}$Rb in arms $A,B$, respectively, of the interferometer configuration in \Cref{fig:epinterforemeter}. The order of Bragg diffraction is indicated by $n$ and $g_{85(87)}$ is the gravitational acceleration experienced by ${}^{85(87)}$Rb, which are positive in our conventions. The difference between the phases of the two species is
\begin{equation}
\Delta \Phi=-n k \Delta g T \, ,
\end{equation}
from which one can extract the relative acceleration $\Delta g=g_{85}-g_{87}$ and consequently the Eötvös parameter defined as
\begin{equation}
    \eta=2\frac{g_{85}-g_{87}}{g_{85}+g_{87}} \, .
\end{equation}

Upon introducing IR/UV mixing effects through the modified dispersion relation \eqref{eq:mdr1alternate1}, the accumulated phase differences $\Delta\phi_A$ and $\Delta\phi_B$ are corrected as follows

\begin{equation}
\begin{aligned}
  \delta (\Delta\phi_A) =& -\ell \bigg({m_{85} T(g_{85} m_{85}T + n  k)}+\\
  &-{m_{87} T(g_{87} m_{87}T + n  k)}\bigg) \, ,
  \end{aligned}
\end{equation}
\begin{equation}
\begin{aligned}
  \delta (\Delta\phi_B) =& -\ell \bigg({m_{85} T(g_{85} m_{85}T - n  k)}+\\
  &-{m_{87} T(g_{87} m_{87}T - n  k)}\bigg) \, .
  \end{aligned}
\end{equation}
The corrections to the phase difference can be recast as a correction to the Eötvös parameter
\begin{equation}
    \eta=\eta_0\bigg(1-4 \ell \frac{m_{85}-m_{87}}{(g_{85}+g_{87})T\eta_0} \bigg) \, .
\end{equation}

\end{document}